\documentclass[suppldata]{interact}

\usepackage{epstopdf}% To incorporate .eps illustrations using PDFLaTeX, etc.
\usepackage{lineno,hyperref,ap,graphics,graphicx,color}
\usepackage[caption=false]{subfig}% Support for small, `sub' figures and tables

\usepackage{natbib}% Citation support using natbib.sty
\bibpunct[, ]{(}{)}{;}{a}{}{,}% Citation support using natbib.sty
\theoremstyle{plain}% Theorem-like structures provided by amsthm.sty

\theoremstyle{definition}

\theoremstyle{remark}

\begin{document}

%\articletype{Research Article}% Specify the article type or omit as appropriate

\title{Concurrent consideration of cortical and cancellous bone within continuum bone remodelling}

\author{
	\name{Ina Schmidt\textsuperscript{a}\thanks{CONTACT Ina~Schmidt Author. Email: ina.schmidt@th-nuernberg.de}, Areti~Papastavrou\textsuperscript{a},  and Paul Steinmann\textsuperscript{b}}
	\affil{\textsuperscript{a}Faculty of Mechanical Engineering, Nuremberg Tech, Ke\ss{}lerplatz 12, 90489  Nuremberg, Germany;  \textsuperscript{b}Chair of Applied Mechanics, University of Erlangen-Nuremberg, Egerlandstra\ss{}e 5, 91058 Erlangen, Germany}
}

\maketitle

\begin{abstract}
Continuum bone remodelling is an important tool for predicting the effects of mechanical stimuli on bone density evolution.
While the modelling of only cancellous bone is considered in many studies based on continuum bone remodelling, this work presents an approach of modelling also cortical bone and the interaction of both bone types.
The distinction between bone types is made by introducing an initial volume fraction.
A simple point-wise example is used to study the behaviour of novel model options, as well as a proximal femur example, where the interaction of both bone types is demonstrated using initial density distributions.
The results of the proposed model options indicate that the consideration of cortical bone remarkably changes the density evolution of cancellous bone, and should therefore not be neglected.
\end{abstract}


%\linenumbers

%%%%%%%%%%%%%%%%%%%%%%%%%%%%%%%%%%%%%%%%%%%%%%%%%%%%
\section{Introduction}

Tubular bone is characterised by cortical bone surrounding the bone marrow in the shaft and the cancellous bone at the proximal and distal ends.
Due to its special microstructure composed of beams and plates (trabeculae), cancellous bone is also denoted as spongy bone.
Depending on environmental changes, such as mechanical loading, bone can adapt its internal microstructure by altering the density and trabecular orientation.
Remodelling of cancellous bone was first formulated by \citep{Wolff1892} and is referred to as Wolff's law of bone remodelling.
Based on continuum mechanics and the finite element method, simulation approaches to bone remodelling were developed, see \citep{Cowin1976}, \citep{Huiskes1987}, \citep{Weinans1992}, \citep{Harrigan1993}, \citep{Jacobs1995} and many more.
Most of these studies address exclusively the adaptation of cancellous bone.
Even if the focus is also on the adaptation of cortical bone, it is typically considered separately from spongy bone, see \citep{Carter2010}.
One of the few approaches involving the interaction of both bone types was investigated by \citep{HAMBLI2014} and \citep{BARKAOUI2019} based on the cellular activities of bone.
Another proposal was developed by \citep{Nutu2018} based on the model of \citep{Huiskes1987} for dental implants.\\

This work aims to simulate bone density changes of cancellous and cortical bone as well as their interplay based on a continuum bone remodelling approach.
Unlike spongy bone, the bone fibrils of cortical bone are densely packed, which is why this lamellar structure is also called compact bone.
Remodelling of cortical bone can occur along the endosteal and periosteal surface as well as within the channels.
The surface area per unit volume is considerably reduced compared to cancellous bone and therefore remodelling requires more time with less change in density.
In order to adapt to the load case, the main mechanism considered here is the adjustment of the microstructural density.
Whereas a change in bone geometry will not be considered, the simulation of bone adjustments in the endosteal area such as trabecularisation and thinning of the inner compact bone in old age is made possible.

Based on previous studies, see \citep{Kuhl2003}, \citep{Liedtke2017} \citep{Papastavrou2020} and \citep{Papastavrou2020b}, an initial volume fraction is introduced in this work, which is defined as the ratio between initial nominal density and true density and therefore allows a distinction between the two bone types.
The influence of this new parameter and the associated modelling of spongy and compact bone will be studied in this context.

In the following, the quantities $\{\circ\}$, which take into account the effective mixture of solid bone material and the (empty) pore space are referred to as \textit{nominal} quantities, whereas \textit{true} quantities based on the properties of the solid (bone) material are denoted as $\{\circ\}^s$.
Furthermore we distinguish between the \textit{initial} value $\{\circ\}^\star$ of a quantity (at time zero), and its \textit{current} (time-evolving) value $\{\circ\}(t)$.
The latter should not be confused with the \textit{actual} configuration, which refers to the geometric description of the deformed continuum body.

This contribution is organised as follows.
The governing and constitutive equations including the proposed modification are detailed in Section \ref{sec2}.
The performance is illustrated by numerical problems in Section \ref{Sec3}.
Section \ref{Sec4} provides a concluding discussion.

%%%%%%%%%%%%%%%%%%%%%%%%%%%%%%%%%%%%%%%%%%%%%%%%%%%%
\section{Continuum framework of bone remodelling} \label{sec2}

With a view on its continuum modelling, bone is here considered as a porous material consisting of a solid skeleton and (empty) pore space, see Fig. \ref{fig:mic_mac}.
In the following, the underlying equations of continuum bone remodelling are briefly re-iterated, see also \citep{Kuhl2003} and \citep{Holzapfel2001} for additional details.

\subsection{Kinematics}

The kinematic description is characterised by the deformation map $\boldsymbol{\varphi}$ that maps the referential placement $\mathbf{X}$ of a physical particle in the material (reference) configuration $\mathcal{B}_{0}$ at time $t_0$ to its actual placement $\mathbf{x}$ in the spatial (actual) configuration $\mathcal{B}_{t} \subset \mathbb{E}^3$ at time $t  \in \mathbb{R}_{+}$

\begin{equation}
\mathbf{x} = \boldsymbol{\varphi}(\mathbf{X},t)  : \mathcal{B}_0 \times \mathbb{R}_{+} \to \mathcal{B}_{t} \:.
\end{equation}

The referential gradient $\nabla_0$ of the nonlinear deformation map denotes the deformation gradient $\mathbf{F}$, it maps from the material tangent space $T \mathcal{B}_{0}$ to the spatial tangent space $T\mathcal{B}_{t}$

\begin{equation}
\mathbf{F} = \nabla_0 \boldsymbol{\varphi}(\mathbf{X},t)  : T\mathcal{B}_0 \to T\mathcal{B}_{t} \:.
\end{equation}

The determinant of the deformation gradient is the Jacobian $J=\det \mathbf{F} >0$ and relates the volume elements between the two configurations.
The right and left Cauchy-Green tensor $\mathbf{C} = \mathbf{F}^T \cdot \mathbf{F}$ and $\mathbf{b} = \mathbf{F} \cdot \mathbf{F}^T$ are introduced as characteristic strain measures.\\

In the following, the expression $\dot{\overline{ \{\circ\} }} = \partial_t \{\circ\} \vert_X $ is used for the material time derivative of a quantity $\{\circ\}$ at fixed material placement $\mathbf{X}$.
The presented formulation refers completely to the material configuration.
Therefore, $\nabla_0 \{\circ\}$ and $\Div \{\circ\}$ denote the (referential) gradient and divergence of any field $\{\circ\}$ with respect to the material placement $\mathbf{X}$.
A density quantity $\{\circ\}$ may be expressed per unit volume $\{\circ\}_0$ or per unit mass $\{\circ\}$, both related via the current referential \textit{nominal} density $\rho_{0}$ as $\{\circ\}_0 = \rho_{0} \{\circ\}$.

\subsection{Balance equations}

The balance of linear momentum is given by 

\begin{equation}
\Div \mathbf{P} = \mathbf{0}
\end{equation}

with $\mathbf{P} = \mathbf{P} (\nabla_0 \boldsymbol{\varphi}, \rho_{0},...) $ the Piola stress tensor.
Since the time scale of inertial effects is significantly smaller than that of the remodelling process, the inertial term is neglected herein \citep{Frost1987}.
The loads applied during locomotion typically correspond to a multiple of gravity. 
Therefore, body forces are usually not considered in bone remodelling as they are not the dominant stimulus \citep{Jacobs1995}.

The balance of mass is given by

\begin{equation}
\dot{\rho_{0}} = R_0
\end{equation}

where $R_0$ is the scalar-value mass source term.
As the actual \textit{nominal} density  $\rho_{t} = {\rho_{0}}/{J}$ is time-dependent also due to the coupling with the Jacobian, it is useful to capture the evolution of the referential \textit{nominal} density $\rho_{0}$.
The relative change in (referential) nominal density $\tilde{\rho_{0}}$ is defined in terms of the \textit{initial} nominal density $\rho_{0}^\star$ as

\begin{equation}
\tilde{\rho_{0}} = \frac{\rho_{0}-\rho_{0}^\star}{\rho_{0}^\star} .
\end{equation}

Therefore, density gain results for values $\tilde{\rho_{0}} >0$ whereas density loss is obtained for $-1 < \tilde{\rho_{0}} < 0$.

\begin{figure}[ht]
	\centering
	\includegraphics[width=0.7\textwidth]{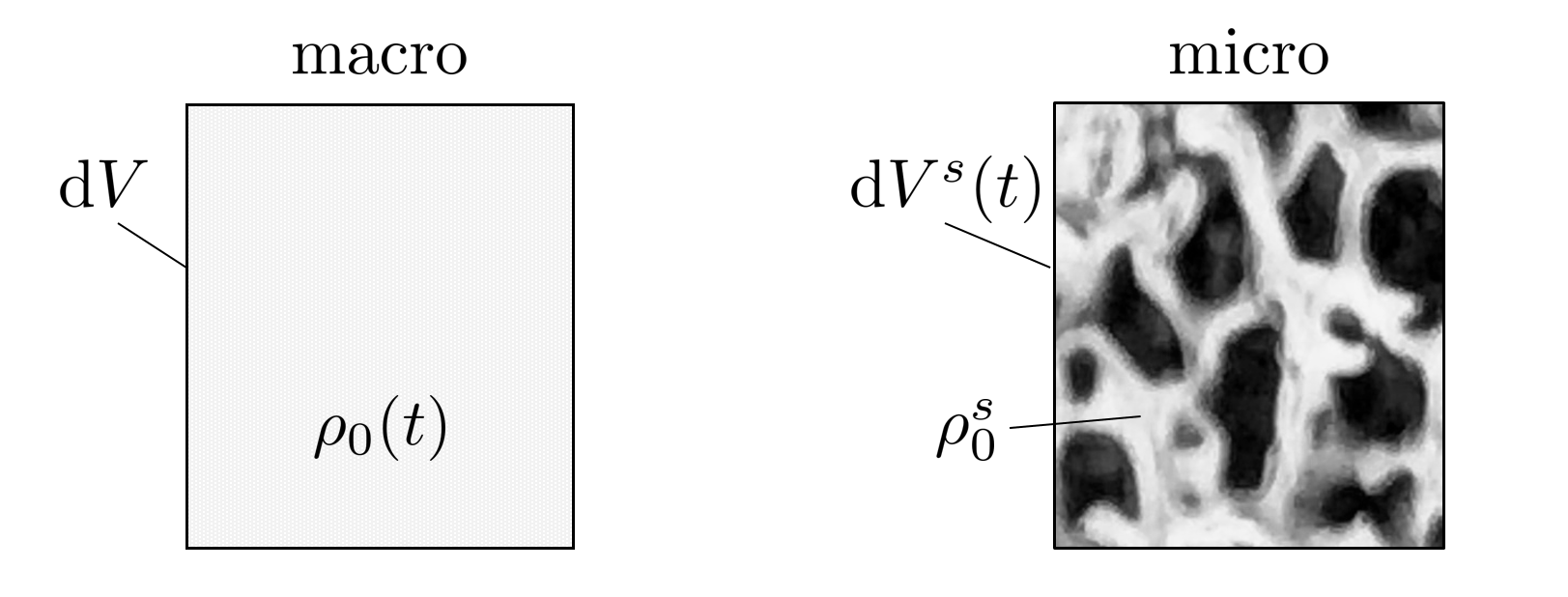}
	
	\caption{Homogenised macro scale with nominal density $\rho_0(t)$ versus heterogeneous micro scale with true density $\rho_0^s$}
	
	\label{fig:mic_mac}
\end{figure}

\subsection{Nominal versus true mass densities}

We distinguish between the homogenised macro scale with \textit{nominal} (time-evolving, referential) density $\rho_0=\rho_0(t)\in(0,\rho_0^s]$ occupying the (time-independent, referential) volume element $\d V$ and the heterogeneous micro scale with \textit{true} (time-independent, referential) density $\rho_0^s$ of the solid skeleton material occupying the (time-evolving, referential) partial volume element $\d V^s=\d V^s(t)\in(0,\d V]$, see Fig. \ref{fig:mic_mac}. The (time-evolving, referential) \textit{volume fraction} of solid skeleton material thus follows as

\begin{equation}
\nu_0=\nu_0(t):=\frac{\d V^s(t)}{\d V}=\frac{\rho_0(t)}{\rho_0^s}\in(0,1].
\end{equation}

The (referential) volume fraction $\nu_0(t)$ indicates how much space within a (referential) volume element $\d V$ at the macro scale is occupied by solid skeleton material, thus $1-\nu_0(t)$ denotes the (referential) volume fraction of the (empty) pore space.\\

At time $t=0$ the corresponding \textit{initial} values for the (referential) volume fraction $\nu_0(0)$, the (referential) partial volume element $\d V^s(0)$ and the nominal (referential) density $\rho_0(0)$ are related via

\begin{equation}
\nu_0^\star:=\frac{\d V^{s\star}}{\d V}=\frac{\rho_0^\star}{\rho_0^s}\in(0,1].
\end{equation}

and are denoted the \textit{initial} (referential) volume fraction $\nu_0^\star:=\nu_0(0)$, the \textit{initial} (referential) partial volume $\d V^{s\star}:=\d V^s(0)$ and the \textit{initial} nominal (referential) density $\rho_0^\star:=\rho_0(0)$, respectively. Typically, $\nu_0^\star$ is significantly higher for cortical bone as compared to cancellous bone.\\

Furthermore, we may distinguish between the \textit{nominal} relative (referential) density and its \textit{true} counterpart

\begin{equation}
\frac{\rho_0(t)}{\rho_0^\star}\equiv\frac{\nu_0(t)}{\nu_0^\star}\in(0,\infty)\quad\mbox{versus}\quad
\frac{\rho_0(t)}{\rho_0^s}\equiv\nu_0(t)\in(0,1]
\end{equation}

when phenomenologically modelling the homogenised mechanical properties of bone based on its skeleton structure at the micro scale, i.e. its \textit{microstructure}. In the following, the present contribution aims to rationalise various modelling options stemming from this distinction.

\subsection{Nominal versus true free energy densities}

Bone possesses a microstructure characterised by the spatially heterogeneous distribution of solid skeleton and pore space. The resulting structural bone properties of the heterogeneous solid skeleton at the micro scale may then be homogenised into \textit{nominal} bone properties of the effective material at the macro scale.\\

A simplistic, however efficient and effective homogenisation rule -- typically employed for cellular materials -- relates the \textit{true} material parameters of the solid skeleton at the micro scale to the corresponding effective, \textit{nominal} parameters of the homogenised material at the macro scale via a power $N$ of the \textit{initial} (referential) volume fraction $\nu_0^\star$. Let for example $\lambda^s$ and $\mu^s$ denote the Lamé parameters of the solid skeleton material. Then, the homogenised, effective bone material is characterised by the \textit{nominal} material parameters

\begin{equation}
\lambda^\star = [\nu_0^\star]^N \lambda^s \quad\mbox{and}\quad
\mu^\star = [\nu_0^\star]^N \mu^s \quad\mbox{with}\quad N=\left\lbrace 1,n \right\rbrace \:.
\end{equation}

Typically, the concrete values assumed for the \textit{homogenisation exponent} $N$ capture different microstructures and homogenisation approaches.
In this contribution, either the most elementary volumetric mixture rule with $N=1$ or a relation with $N=n$ shall be considered.
The latter is particularly motivated by the proposal of \citep{Gibson2005} postulating a dependence of the \textit{nominal} Young's modulus of a homogenised (linear elastic) porous material on the Young's modulus of the solid skeleton material via a power of the \textit{true} relative density.\\

In what follows we assume the concept of \textit{strain equivalence} as also employed in damage mechanics. To elucidate this aspect more concretely, we employ, without loss of generality, a (referential) volume-specific Neo-Hookean \textit{true} free energy density for the solid skeleton material

\begin{equation}
\Psi_0^s = \frac{1}{2} \lambda^s \ln^2 J +\frac{1}{2} \mu^s \left[ \mathbf{F} : \mathbf{F} -3-2\ln J \right] .
\end{equation}

Note that strain equivalence postulates identical 'strain' at the macro and the micro scale. 
Thus, based on the \textit{nominal} material parameters, the corresponding (referential) volume- specific initial \textit{nominal} free energy follows from its \textit{true} counterpart as

\begin{equation}
\Psi_0^\star  =  [\nu_0^\star]^N \Psi_0^s\:.
\end{equation}

Concretely, the (referential) volume-specific Neo-Hookean initial \textit{nominal} free energy reads in terms of the \textit{nominal} Lamé parameters as

\begin{equation}
\Psi_0^\star = \frac{1}{2} \lambda^\star \ln^2 J +\frac{1}{2} \mu^\star \left[ \mathbf{F} : \mathbf{F} -3-2\ln J \right] \:.
\end{equation}

The \textit{initial} microstructure evolves with time due to remodelling into the \textit{current} microstructure, thus necessitating to express the (referential) volume-specific \textit{current} nominal free energy density $\Psi_0$ in terms of its \textit{initial} counterpart.

To this end, again motivated by \citep{Gibson2005}, when taking the microstructure of cellular materials into account, the (referential) volume-specific \textit{current} nominal free energy density $\Psi_0$ is determined by weighting its \textit{initial} nominal counterpart $\Psi_0^\star$ with a power $n$ of the \textit{nominal} relative density

\begin{equation}
\Psi_0 = \left[ \frac{\rho_{0}}{\rho_{0}^\star}\right]^n \Psi_0^\star \:.
\end{equation}

Then the Piola stress entering the equilibrium equation derives from the (referential) volume-specific \textit{current} nominal free energy density $\Psi_0$ as

\begin{equation}
\mathbf{P}:=\frac{\partial\Psi_0}{\partial\mathbf{F}}=\left[ \frac{\rho_0}{\rho_0^\star}\right]^n\frac{\partial\Psi_0^\star}{\partial\mathbf{F}}=:
\left[ \frac{\rho_0}{\rho_0^\star}\right]^n\mathbf{P}^\star
\end{equation}

and expands concretely for the Neo-Hookean case with the \textit{initial} Piola stress

\begin{equation}
\mathbf{P}^\star=\lambda^\star\ln J\,\mathbf{F}^{-t}+\mu^\star[\mathbf{F}-\mathbf{F}^{-t}]\:.
\end{equation}

Incorporating the (referential) volume-specific Neo-Hookean \textit{true} free energy density $\Psi_0^s$ eventually renders $\Psi_0$ in terms of the power $n$ of the \textit{true} relative density and the power $N-n$ of the \textit{initial} (referential) volume fraction

\begin{equation}
\Psi_0 
= \left[ \frac{\rho_0}{\rho_0^s}\right]^n [\nu_0^\star]^{N-n}\, \Psi_0^s
= [\nu_0]^n [\nu_0^\star]^{N-n}\, \Psi_0^s\:.
\end{equation}

The exponent $n$ is typically set to $n=2$ for trabecular bone with low nominal density and to $n> 2$ for denser cancellous bone, compare to \citep{Rho1993} and \citep{Gibson2005}.\\

Setting finally the homogenisation exponent $N$ in particular to $N=n$ results in an expression for the (referential) volume-specific \textit{current} nominal free energy density $\Psi_0$ in terms of its \textit{true} counterpart and the \textit{true} relative density

\begin{equation}
\Psi_0 \stackrel{N=n}{=} \left[ \frac{\rho_0}{\rho_0^s}\right]^n \Psi_0^s 
=[\nu_0]^n \Psi_0^s\:.
\end{equation}

Correspondingly, the Piola stress expands in this case as

\begin{equation}
\mathbf{P}\stackrel{N=n}{=}\left[ \frac{\rho_0}{\rho_0^s}\right]^n\frac{\partial\Psi_0^s}{\partial\mathbf{F}}=:
\left[ \frac{\rho_0}{\rho_0^s}\right]^n\big[\lambda^s\ln J\,\mathbf{F}^{-t}+\mu^s[\mathbf{F}-\mathbf{F}^{-t}]\big]\:.
\end{equation}

Setting $N=n$ tacitly assumes that the geometry of the microstructure is not distorted by the deformation (for example a circular pore remains a circular pore). In contrast, setting $N\not=n$ conceptually allows to distinguish between the \textit{initial} and the (deformation-dependent) \textit{current} microstructure, thus in general $n=n(\mathbf{F})$ with $N=n(\mathbf{I})$.\\

This contribution shall explore the consequences of two limiting cases by setting $N$ to either $N=1$ (volumetric mixture rule) or to $N=n$ (cellular material like homogenisation rule). More sophisticated scenarios with $n=n(\mathbf{F})$ where $N=n(\mathbf{I})$ shall not be analysed here.

\subsection{Mass density flux}

To model changes in the \textit{nominal} (referential) density $\rho_0$ incurred by mechanical stimuli, the volume-localised mass balance $\dot\rho_0=R_0$ is equipped with a (referential) source term $R_0$. Following \citep{Harrigan1993}, the (referential) mass source $R_0$ is expressed by the (referential) volume-specific \textit{current} nominal free energy density $\Psi_0$ and the  \textit{nominal} relative density as

\begin{equation} \label{eq20}
R_{0} = c \left[ \left[ \frac{\rho_{0}}{\rho_{0}^\star} \right]^{-m} \, \Psi_{0} - \Psi_{0}^a \right] \:.
\end{equation}

Here, the parameter $c$ controls the speed of density evolution and has dimension of time divided by length squared.
The attractor stimulus $\Psi_{0}^a$ defines the biological equilibrium (homeostasis) to which the system is driving at constant (mechanical) stimulus.
The exponent $m$ is  a material parameter particularising the mechanical stimulus \citep{Harrigan1993} in terms of the \textit{nominal} relative density.\\

The (referential) mass source may alternatively be expressed by the (referential) volume-specific \textit{initial} nominal free energy density $\Psi_0^\star$ or, likewise, by its \textit{true} counterpart $\Psi_0^s$ as

\begin{equation}
R_{0} =  c \left[ \left[ \frac{\rho_{0}}{\rho_{0}^\star} \right]^{n-m} \, \Psi_{0}^\star - \Psi_{0}^a \right] 
=  c \left[ \left[ \frac{\rho_{0}}{\rho_{0}^s} \right]^{n-m} [\nu_0^\star]^{N-n+m}\, \Psi_{0}^s - \Psi_{0}^a \right] \:.
\end{equation}

We may deduce from this representation that the exponent $m$ is also crucial for the stability of the density evolution: at homeostasis ($R_0=0$) $\rho_{0}>\rho_{0}^\star$ for $\Psi_{0}^\star > \Psi_{0}^a$ only if $m-n>0$. In conclusion, the exponent $m$ should be chosen as $m>n$ \citep{Harrigan1994}, \citep{Kuhl2003}.\\

Setting finally the homogenisation exponent $N$ in particular to $N=n$ results in alternative expressions for the (referential) mass source

\begin{equation}
R_{0} =  c \left[ \left[ \frac{\rho_{0}}{\rho_{0}^s} \right]^{n-m} [\nu_0^\star]^m\, \Psi_{0}^s - \Psi_{0}^a \right]
=  c \left[ \left[ \frac{\rho_{0}}{\rho_{0}^\star} \right]^{n-m} [\nu_0^\star]^n\, \Psi_{0}^s - \Psi_{0}^a \right] \:.
\end{equation} 

%%%%%%%%%%%%%%%%%%%%%%%%%%%%%%%%%%%%%%%%%%%%%%%%%%%%
\section{Simulation and results} \label{Sec3}

The performance of the proposed model options ($N=n$ versus $N=1$) is demonstrated using classical benchmark problems.
The first example considers a unit square specimen as defined by \citep{Kuhl2003} to study temporal bone density evolution at a single point.
Second, we analyse the consequence of the spatial distribution of bone microstructure using a two-dimensional proximal femur head as given by \citep{Carter2010}.

\subsection{Investigating the influence of the initial volume fraction} \label{square}

\begin{figure}[ht]
	\centering
	\includegraphics[width=0.95\textwidth]{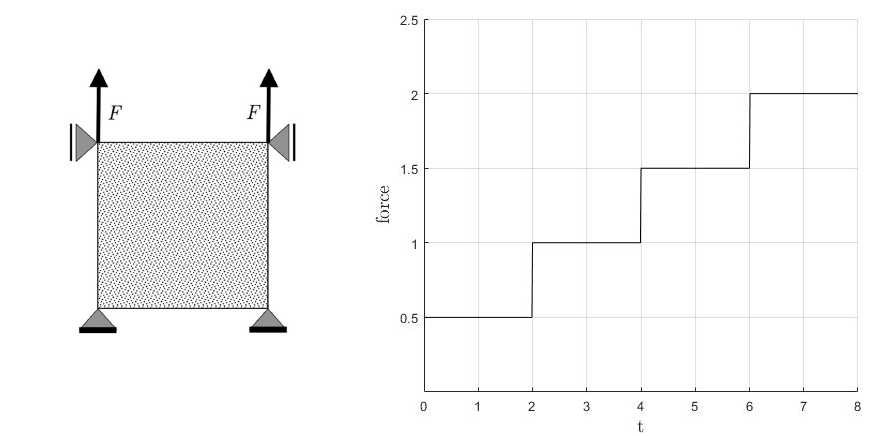}
	
	\caption{Unit square example: specimen and load function}
	
	\label{fig:square}
\end{figure}

To differentiate between cancellous and cortical bone tissue, the effect of the initial volume fraction shall be examined.
For this purpose, a simple unit square sample is studied, which is loaded in tension by a force function that increases stepwise as illustrated in Figure \ref{fig:square}.

The reference parameters are chosen similar to \citep{Kuhl2003}: \textit{true} elastic modulus $E^s=1.0$, Poisson's ratio $\nu^s=0.0$, attractor stimulus $\Psi_{0}^a=1.0$, parameter $m=3$ and parameter $n=2$.
The homogenisation exponent is chosen as $N=n$ as motivated by the proposal of \citep{Gibson2005}.\\

A \textit{true} density of the solid material $\rho_{0}^s=2 $, similar to \citep{Gibson2005}, is assumed for both bone microstructure types in the following.
It is known from measurements that the \textit{true} density of trabeculae is slightly smaller and the one of cortical bone fibrils slightly larger, see \citep{An2000}.
However, as the objective of this analysis is to point out the general trend in the evolution of the different tissue microstructures, the assumption of equal true density is sufficient in this context.
An initial \textit{nominal} density of $\rho_{0}^\star = 0.5 $ leading to an initial volume fraction of $\nu_0^\star= {\rho_{0}^\star}/{\rho_{0}^s}= {0.5}/{2}=0.25$ is chosen as representative of cancellous bone.
A comparatively high value of $\nu_0^\star= {\rho_{0}^\star}/{\rho_{0}^s}={1.8}/{2}=0.9$ is set for the volume fraction representing compact bone.

\begin{figure}[ht]
	\centering
	\includegraphics[width=1\textwidth]{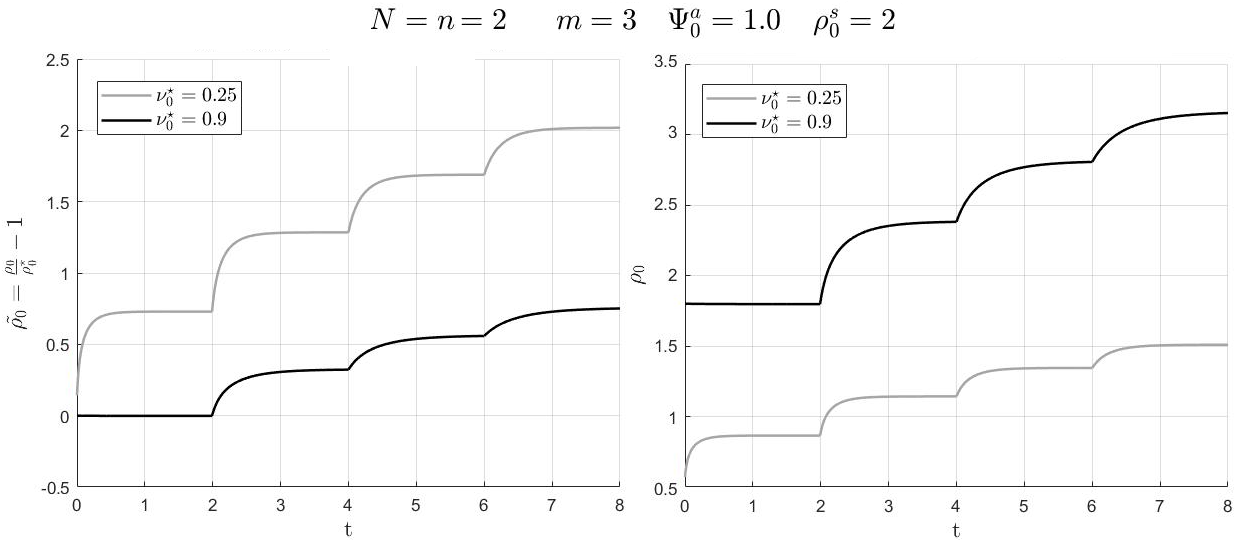}
	
	\caption{Unit square example: evolution of relative change in nominal density and absolute nominal density with a variation in the initial volume fraction representing cortical and spongy bone}
	
	\label{fig:sq_ks}
\end{figure}

As shown in Figure \ref{fig:sq_ks}, a small initial volume fraction (gray line) leads to rapid bone growth while a high value (black line) causes no density change in the first load step.
Only after increasing the load in the second step, and thus increasing the mechanical stimulus, an adaptation of bone density can be observed.\\

Cortical bone in the model therefore requires a significantly higher stimulus and grows less in relative terms.
But despite its rapid growth, the absolute density of cancellous bone still remains at a lower level.

\subsection{Investigating the effect of exponent $m$ and the attractor stimulus} \label{sec_m}

Due to its small surface area in relation to the volume, cortical bone adapts to the load conditions much slower and less than spongy bone.
Considering the mass source term in Equation \ref{eq20}, it appears that there are three parameters which can influence the behaviour of bone adaptation.
By reducing the parameter $c$, the speed of density evolution can be slowed down, but after a sufficiently long time the result always converges towards the same saturation level.
Therefore, a parameter adjustment of the exponent $m$ and the attractor stimulus $\Psi_{0}^a$ shall be considered for cortical bone with $\nu_0^\star=0.9$ and $N=n=2$.\\

\begin{figure}[ht]
	\centering
	\includegraphics[width=1\textwidth]{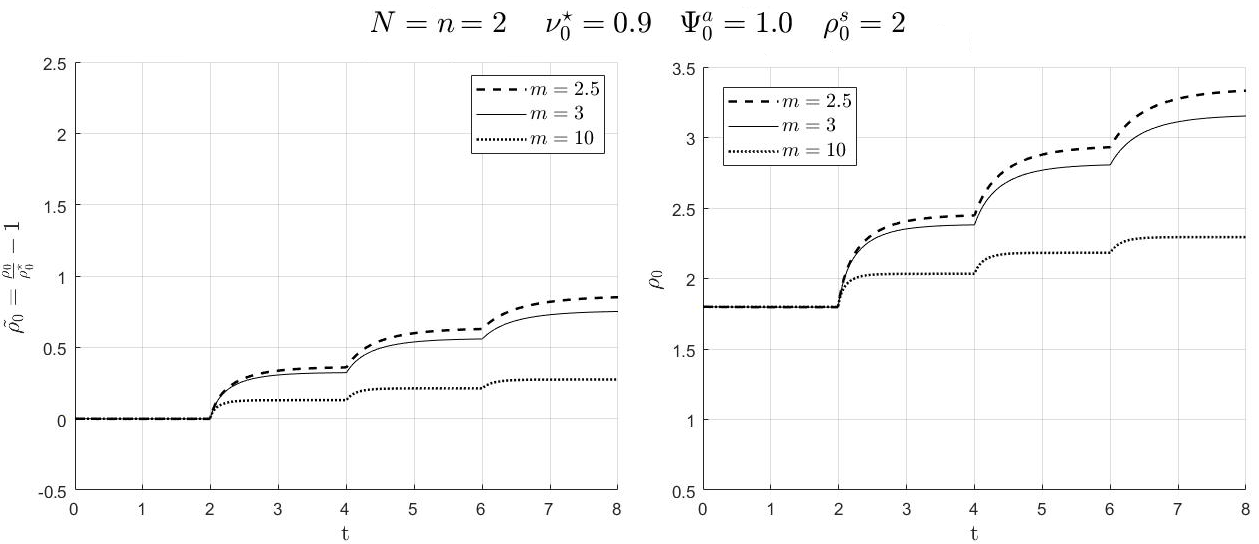}
	
	\caption{Unit square example: evolution of relative change in nominal density and absolute nominal density with variations of $m$ for cortical bone}
	
	\label{fig:sq_m}
\end{figure}

In order to satisfy the stability criterion, the exponent $m$ must be selected greater than $n$, see \citep{Harrigan1993}.
Thus, $m=3$, $m=2.5$ and $m=10$ are investigated as depicted in Figure \ref{fig:sq_m}.

As can be seen, reducing the exponent leads to an increase in density growth (dashed line).
In contrast, choosing a large value for the exponent $m$ tends to slow down the density evolution (dotted line).

One possibility to simulate the significantly reduced density evolution of compact bone compared to spongy bone would therefore be setting a relatively high exponent $m$.\\

\begin{figure}[h]
	\centering
	\includegraphics[width=1\textwidth]{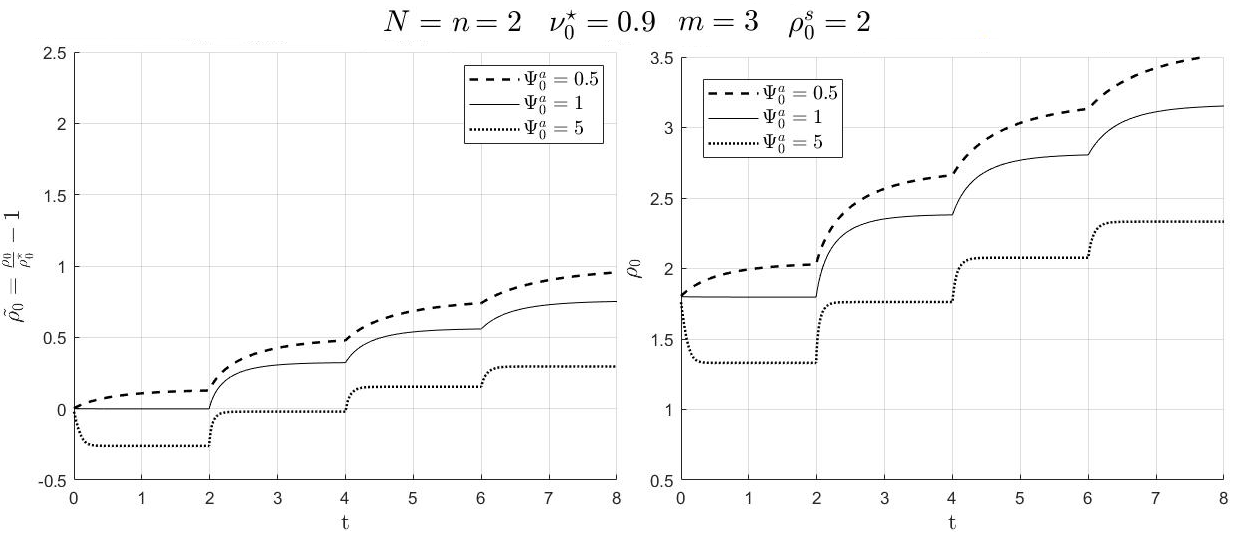}
	
	\caption{Unit square example: evolution of relative change in nominal density and absolute nominal density with variations of the attractor stimulus for cortical bone}
	
	\label{fig:sq_psi}
\end{figure}

A variation of the attractor stimulus is analysed in Figure \ref{fig:sq_psi}.

The dotted line in Figure \ref{fig:sq_psi} based on a relatively high attractor stimulus displays low density results.
Bone growth can not be achieved until the third load level.
Reducing the attractor stimulus (dashed line) results in an overall increased gain in nominal density.\\

Since cortical bone should neither grow nor absorb faster than cancellous bone, in the following, the same attractor stimulus will be assumed for both microstructure types.

\subsection{Investigating the interaction between cancellous and cortical bone} \label{sec_fem}

To analyse the local distribution of cortical and cancellous bone and their interaction, a two-dimensional proximal femur head is examined.

\begin{figure}[ht]
	\centering
	\includegraphics[width=1\textwidth]{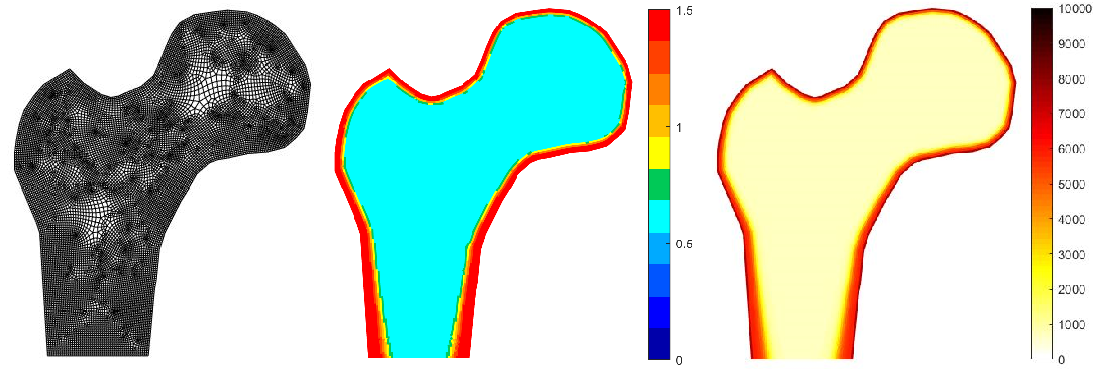}
	
	\caption{Femur example: mesh and the distribution of the initial nominal density and the nominal Young's modulus}
	
	\label{fig:ini_d}
\end{figure}

Three typical load cases representing the midstance phase of gait, extreme abduction and adduction are applied to the model as reported in \citep{Carter2010}.
In the calculations bi-linear element expansions with a 2x2 Gauss quadrature rule are used.
The mesh, see Figure \ref{fig:ini_d} (left), consists of 13317 elements and 13602 nodes.
The parameters are set as follows: Poisson's ratio $\nu^s=0.2$, attractor stimulus $\Psi_{0}^a = 0.01$, exponent $n=2$ and true solid material density $\rho_{0}^s = 2.0$.

An initial \textit{nominal} cortical density of $\rho_{0}^\star=1.825$, leading to an initial volume fraction of $\nu_0^\star= {\rho_{0}^\star}/{\rho_{0}^s}= {1.825}/{2}=0.9125$ is set for cortical bone (elements at the external boundary),
and $\rho_{0}^\star=0.55$ with $\nu_0^\star= {\rho_{0}^\star}/{\rho_{0}^s}= {0.55}/{2}=0.275$ for cancellous bone.
To avoid an abrupt unrealistic change between elements of different microstructure types, a small transition zone between both regions is created using a sigmoid function, see Figure \ref{fig:ini_d} (middle).\\

Four different models are studied and compared in the following, all using the homogenisation exponent $N=n=2$.
Selected parameters and thus the difference between the four models are illustrated in Table \ref{table:femur}.

\begin{table}[h]
	\tbl{non-dimensionalised reference parameters of the four proxima femur models with $N=n=2$}
	{\begin{tabular}{l|cccc} \toprule
			   & ${\nu_0^\star}$ of cancellous bone & ${\nu_0^\star}$ of cortical bone & $m$ of cancellous bone & $m$ of cortical bone \\
			  \toprule
			Model 1 & 0.275 & --- & 3 & ---\\  
			Model 2 & 0.275 & 0.9125 & 3 & 3 \\
			Model 3 & 0.275 & 0.9125 & 3 & 10 \\
			Model 4 & 0.275 & 0.9125 & 3 & 18 \\
			\hline
	\end{tabular}}
	\label{table:femur}
\end{table}

The first model represents the full spongy bone as reported in previous literature with an initial \textit{nominal} density of $\rho_{0}^\star=0.55$.
The second model differentiates between cortical and cancellous bone using the initial volume fraction or rather the initial \textit{nominal} density distribution as depicted in Figure \ref{fig:ini_d} (middle).\\

With a \textit{true} elastic modulus of $E^{s}=10000$, which is similarly documented for solid trabecular material by \citep{Rho1993} and \citep{Yamada2014}, and based on the quadratic relation $N=n=2$ between stiffness and \textit{nominal} density according to \citep{Gibson2005}, this results in a \textit{nominal} stiffness of $E^\star=756.25$ in the spongy region at an initial \textit{nominal} density of $\rho_{0}^\star=0.55$, which roughly reflects the magnitude of stiffness as reported in \citep{Wirtz2000} and which is also utilised in the full spongy model.
The maximum \textit{nominal} elastic modulus of cortical bone is approximately $E^\star=8327$ and thus significantly higher, see Figure \ref{fig:ini_d} (right).\\

\begin{figure}[ht]
	\centering
	\includegraphics[width=1\textwidth]{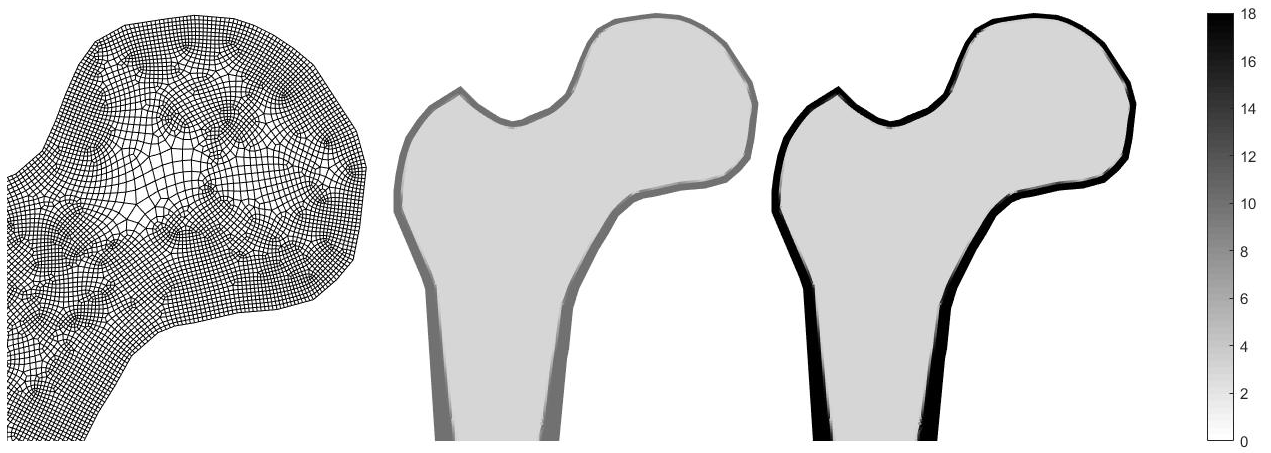}
	
	\caption{Femur example: mesh zoom of femur head and distribution of exponent $m$ with $m=10$ and $m=18$ for cortical bone and $m=3$ for cancellous bone}
	
	\label{fig:ini_m}
\end{figure}

Motivated by the results of section \ref{sec_m}, the third and fourth model include a variation of the exponent $m$.
While the exponent is always set to $m=3$ in the first and second model, the others include an increase within the cortical bone area to a maximum of $m=10$ in the third and $m=18$ in the fourth model.
The distribution of the exponent $m$ follows the respective volume fraction which indicates the type of bone microstructure.
A slight transition is implemented between the two bone types using an hyperbolic tangent function, see Figure \ref{fig:ini_m}.

\begin{figure}[ht]
	\centering
	\includegraphics[width=1\textwidth]{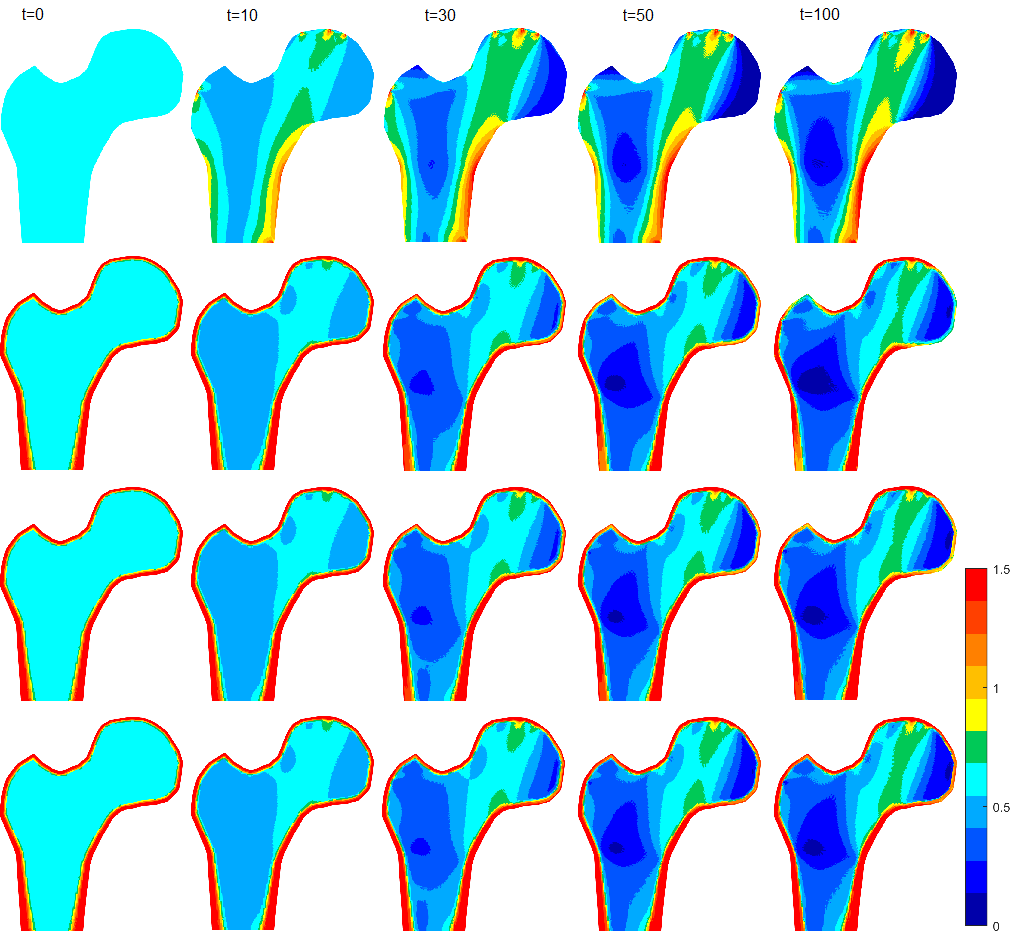}
	
	\caption{Femur example with $N=n=2$: temporal evolution of absolute nominal density of the full spongy model (top row) and with the differentiation of cancellous and cortical bone (second row) and a variation in $m$ for cortical bone: $m=10$ third row, $m=18$ last row}
	
	\label{fig:fem_n}
\end{figure}

The results for the evolution of the \textit{nominal} density are displayed for each model at different points in time in Figure \ref{fig:fem_n}.

Comparing the first two rows, it becomes clear that the pattern of density evolution has changed remarkably.
The density growth in the femoral head and neck is significantly less intense when taking cortical bone  into consideration.
Density loss increases and slightly shifts in the intertrochanter region, which leads to an obvious change in density pattern.
In areas with subliminal stress like the medial femur head, cortical bone decomposes slightly and at the same time ensures a less distinct density reduction in the adjacent spongy region.\\

Increasing the exponent $m$ in cortical bone (see row 3,4) suppresses its resorption and cortical bone remains, although weaker and thinner.
Density loss of cancellous bone in the intertrochanter region is reduced and an increase in density in the femoral neck, albeit small, can be observed compared to the second model.\\

In summary, it appears that cortical bone acts as a kind of stiffer shell.
Therefore, the interior of the femur head bears less load and cancellous bone does not adapt its density as much as in the first, full spongy model.

\subsection{Investigating the impact of the homogenisation exponent N}

\begin{figure}[ht]
	\centering
	\includegraphics[width=1\textwidth]{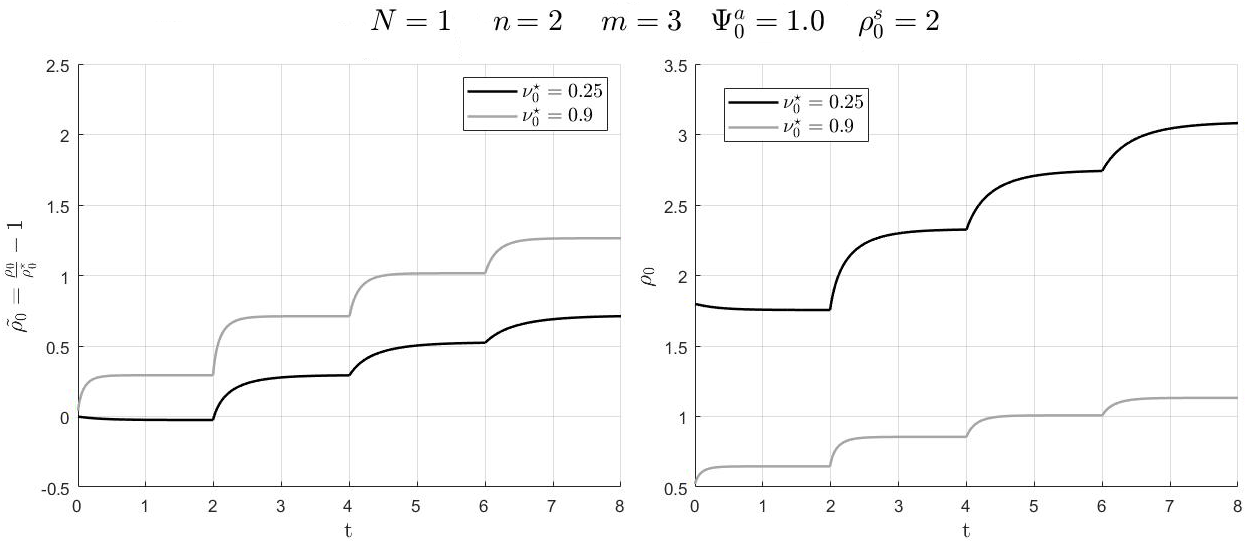}
	
	\caption{Unit square example: evolution of relative change in nominal density and absolute nominal density with N=1 and a variation in the initial volume fraction representing cortical and spongy bone}
	
	\label{fig:sq_ks1}
\end{figure}

Finally, the alternative volumetric mixture rule with $N=1$ shall be examined.
We first re-analyse the simple square example as described in section \ref{square} with a variation in the initial volume fraction and homogenisation exponent $N=1$, whereby all the other parameters remain unchanged.
The temporal evolution of the \textit{nominal} density is shown in Figure \ref{fig:sq_ks1}.

Comparing the results of $N=1$ with those of section \ref{square} in Figure \ref{fig:sq_ks}, it can be seen that density growth is reduced for an initial volume fraction of $\nu_0^\star=0.25$ (gray lines), whereas a reduction in density growth for an initial volume fraction of $\nu_0^\star=0.9$, which even leads to a small density loss in the first load step, is hardly visible (black lines).\\

The reason for this difference lies in the modified relationship between stiffness and \textit{nominal} density from quadratic to linear.
Whereas the initial \textit{nominal} stiffness has quadrupled ($1/0.25$) at an initial volume fraction of $\nu_0^\star=0.25$, the initial volume fraction of $\nu_0^\star=0.9$ yields only a factor of $1.111 = 1/0.9$.\\

\begin{figure}[ht]
	\centering
	\includegraphics[width=0.7\textwidth]{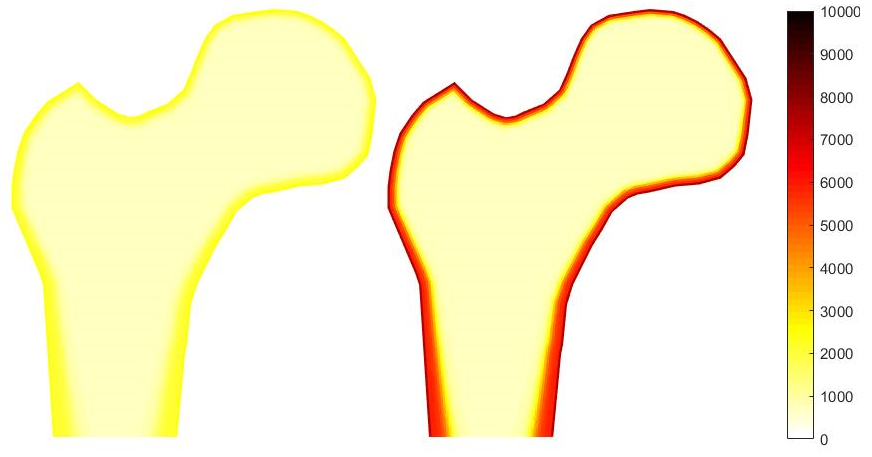}
	
	\caption{Femur example: distribution of nominal Young's modulus with exponent $N=1$ and $E^s=2750$ (left) versus $N=n=2$ and $E^s=10000$ (right)}
	
	\label{fig:ini_E}
\end{figure}

Keeping this in mind, we again consider the proximal femur model, similar to section \ref{sec_fem}.
Setting the \textit{true} elastic modulus to $E^s=10000$ yields an initial \textit{nominal} stiffness of $E^\star=2750$ for spongy bone with $\nu_0^\star=0.275$ via the linear relationship, which is far too high.
Reducing the Young's modulus of the solid material to $E^s=2750$ results in the same $E^\star$ for cancellous bone as in section \ref{sec_fem}, see Figure \ref{fig:ini_E}.
However, the stiffness for cortical bone becomes much too small.\\

\begin{figure}[ht]
	\centering
	\includegraphics[width=1\textwidth]{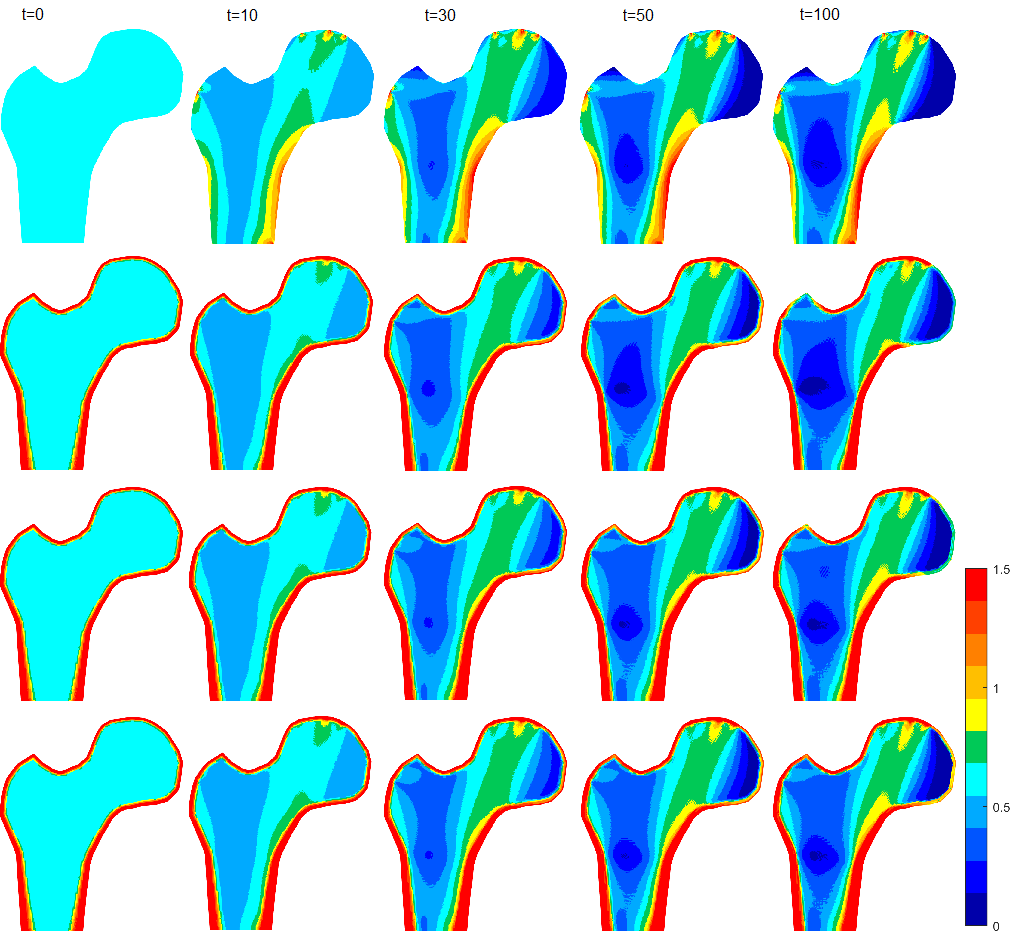}
	
	\caption{Femur example with $N=1$: temporal evolution of absolute nominal density without (top row) and with cortical bone and a variation in $m$ for cortical bone: $m=3$ second row, $m=10$ third row, $m=18$ last row}
	
	\label{fig:fem_1}
\end{figure}

Nevertheless, Figure \ref{fig:fem_1} shows the results of the four different models including the full spongy model, the distinction between cancellous and cortical bone with $m=3$, and two variations of the exponent $m$ for cortical bone, all calculated with a \textit{true} elastic modulus of $E^s=2750$ and an homogenisation exponent of $N=1$.
Since the \textit{true} Young's modulus has been properly adjusted and thus the \textit{nominal} stiffness as well as all the other parameters in the full spongy model correspond to those of section \ref{sec_fem}, the result is the same as can be seen in the first rows of Figure \ref{fig:fem_1} and \ref{fig:fem_n}.\\

When comparing the different homogenisation exponents for the second model, we note the significantly higher density growth in the area of the femoral neck, but also an increased bone loss in the area of the medial femoral head.
Furthermore, the pattern of density evolution more closely resembles that of the full spongy model, especially in the area of the greater trochanter.
This effect is enhanced when increasing the exponent $m$ for cortical bone, see Figure \ref{fig:fem_1} row 3 and 4.
In this scenario, cortical bone decomposes much more in subliminally loaded regions with $N=1$, especially with a small exponent $m$.\\

Overall, the application of the volumetric mixture rule with a homogenisation exponent of $N=1$ leads to a \textit{nominal} density distribution that is more similar to the full-spongy model.
However, ensuring a stiffness-density ratio for cancellous bone as motivated by literature will not simultaneously achieve a realistic ratio in the cortical area.
A further possible option would therefore be a separate definition of an individual \textit{true} Young's modulus for cortical bone, but will not be examined in this survey.

%%%%%%%%%%%%%%%%%%%%%%%%%%%%%%%%%%%%%%%%%%%%%%%%%%%%
\section{Discussion and conclusion} \label{Sec4}

In this study the classical open-system model for bone remodelling by \citep{Kuhl2003} is modified to account for the initial \textit{nominal} density distribution especially at the end of long bones, where two microstructure types, the spongiosa and the compacta can be found.
A distinction is made between the two bone microstructure types by means of the initial volume fraction under the precondition that the same true bone material is assumed for both types.
The model was applied to a one-dimensional benchmark problem and the two-dimensional proximal femur head, which is a representative biomechanical problem.
We demonstrated that the additional modelling of cortical bone influences the density evolution of cancellous bone and should therefore be considered in future patient-specific simulations.

It should be emphasised that this is a first approach to take the local initial density distribution, the remodelling of cancellous and cortical bone and their interaction into account.
Still, realistic material parameters have to be defined via analyses of clinical data, as well as more exact modelling of the boundary conditions and the influence of muscles tendons and ligaments in everyday and extreme loading scenarios.

\section*{Disclosure statement}

None of the authors has a financial or personal relationship with other people or organizations that could 
inappropriately bias his/her work.

\bibliographystyle{tfcse}
\bibliography{bibfile}

\newpage

\listoffigures

\end{document}